# Operando real-space imaging of a structural phase transformation in a high-voltage electrode

*Yifei Sun[1], Sunny Hy[2], Nelson Hua[3,4], James Wingert[3], Ross Harder[5], Ying Shirley Meng[2,6], Oleg Shpyrko[3], Andrej Singer[1,\*]*

[1]*Department of Materials Science and Engineering, Cornell University, Ithaca, New York 14850, USA*
[2]*Department of Nanoengineering, University of California San Diego, La Jolla, California, 92093, USA*
[3]*Department of Physics, University of California San Diego, La Jolla, California, 92093, USA*
[4]*Laboratory for X-ray Nanoscience and Technologies, Paul Scherrer Institut, 5232 Villigen, Switzerland*
[5]*Advanced Photon Source, Argonne National Laboratory, Argonne, Illinois, 60439, USA*
[6]*Pritzker School of Molecular Engineering, University of Chicago, Chicago, Illinois, 60637, USA*
[\*]asinger@cornell.edu

**Abstract**
Discontinuous solid-solid phase transformations play a pivotal role in determining properties of rechargeable battery electrodes. By leveraging *operando* Bragg Coherent Diffractive Imaging (BCDI), we investigate the discontinuous phase transformation in $Li_xNi_{0.5}Mn_{1.5}O_4$ within a fully operational battery. Throughout Li-intercalation, we directly observe the nucleation and growth of the Li-rich phase within the initially charged Li-poor phase in a 500 nm particle. Supported by the microelasticity model, the *operando* imaging unveils an evolution from a curved coherent to planar semi-coherent interface driven by dislocation dynamics. We hypothesize these dislocations exhibit a glissile motion that facilitates interface migration without diffusion of host ions, leaving the particle defect-free post-transformation. Our data indicates negligible kinetic limitations impacting the transformation kinetics, even at discharge rates as fast as C/2. This study underscores BCDI's capability to provide *operando* insights into nanoscale phase transformations, offering valuable guidance for electrochemical materials design and optimization.

**Introduction**
Over a century ago, Gibbs classified phase transformations into two fundamentally different types based on variations in the order parameter[1]. Continuous transformations exhibit variations that are "small in degree but may be great in its extent in space". The initial phase is unstable to infinitesimal fluctuations, resulting in a continuous change of the order parameter across large regions, which can be monitored by measuring macroscopic properties. For example, the order-disorder phase transformation in alloys is studied by measuring diffraction averaged over a large volume[2]. In contrast, discontinuous transformations display variations that are "initially smaller in extent but great in degree"[1]. An energy barrier stabilizes the system against infinitesimally small fluctuations until nucleation occurs at a localized region, causing a disruptive change in the order parameter. A classic example is the martensitic transformation in steel[3]. Capturing intermediate stages of a discontinuous phase transformation is significantly more challenging, as it requires time- and spatially-resolved measurements to observe the nascent nucleus (often only a few nanometers large) and the subsequent interface propagation during growth. Yet, discontinuous phase transformations are of utmost importance in materials science because properties can be tuned by balancing nucleation and growth to achieve the desired microstructure[4,5].

Many electrode materials in rechargeable batteries undergo discontinuous solid-solid phase transformations[6]. These are driven by extraction and insertion of ionic species during charge and discharge. The discontinuous phase transformations are deemed undesirable because the internal stresses at the migrating interface can impede kinetics and lead to mechanical degradations[7]. Nevertheless, many durable, high-rate cathode materials experience phase separation from the discontinuous phase transformation during cycling[8–10]. One example is the disordered spinel phase of $Li_xNi_{0.5}Mn_{1.5}O_4$ (0<x<1) (LNMO), which recently regained significant attention as a cathode material for electric cars because it is cobalt-free and high-voltage. The material shows great potential to outperform its commercial opponents in terms of stability and cost. Additionally, the low nickel content in $Li_xNi_{0.5}Mn_{1.5}O_4$ is particularly appealing as nickel raises environmental and cost concerns similar to cobalt[11,12].

Understanding the mechanisms underpinning phase separation in electrode materials and other functional materials under real operating conditions has become critical of today's materials science: it affects materials systems ranging from quantum materials to energy storage and conversion materials. Addressing this issue requires a three-fold approach. (1) *Operando* measurements with setup that resembles relevant conditions in a working device to capture the transformation. (2) Adequately high temporal resolution to resolve intermediates stages of the transformation. (3) Sufficiently fine spatial resolution in three dimension to distinguish the spatial extent of the nucleated phase and its interface with respect to the parent phase. It can be further complicated by the nano-sized morphology of intercalation systems, which is a way to combat slow diffusion of ions through the host structure[5].

X-ray powder diffraction is a pillar for *operando* characterization of phase transformations in electrochemical systems with sufficient time resolution. During a phase transformation, the diffraction condition differs between the coexisting phases, and each crystalline phase produces a distinct Debye-Scherrer ring[13–15]. Yet powder diffraction averages across many particles within the battery and lacks the spatial resolution to resolve the local structure of the nucleus and the interface during subsequent growth. Recent advances in high-brilliance synchrotron sources have enabled *operando* measurements to capture diffraction signals from individual sub-micron particles[16]. Furthermore, Bragg Coherent Diffractive Imaging (BCDI) provides detailed 3D structural information by inverting coherent diffraction data to real space via an iterative phase retrieval algorithm[17]. This technique measures the strain distribution and buried defects within individual particles[18–21]. In principle, BCDI offers an unparalleled opportunity to image nucleation and growth in individual electrode particles under operating conditions, all without the need for specialized sampling environments. Nevertheless, the challenge of inverting diffraction data comprised of separated diffraction peaks has hindered the realization of real space imaging of phase transformations[22,23].

Here, we use *operando* BCDI to image a discontinuous phase transformation induced by Li-intercalation during galvanostatic battery discharge. In a single 500 nm large $Li_xNi_{0.5}Mn_{1.5}O_4$ particle embedded in a fully operational cell, we directly observe the nucleation and growth of the Li-rich phase inside the initial, fully charged Li-poor phase. *Operando* imaging also reveals the transformation of a curved coherent interface into a planar semi-coherent interface, driven by the introduction of dislocations. These dislocations move with the interface and exit the particle after the phase transformation completes, leaving a defect free crystal. Supported by the microelasticity model, our data show that the dislocation array at the semi-coherent interface reorients the interface during operation. Our data is consistent with a conservative motion of dislocations at the interface.

Consequently, we see no evidence of structural degradation or kinetic limitation from the semi-coherent interface on the phase transformation, even at discharge rates as fast as C/2. Our study highlights BCDI's potential as a robust tool for *operando* insights into structural phase transformations in nanoscale systems, whether induced by electrochemical processes, temperature fluctuations, optical light exposure, or electronic excitation.

**Results**

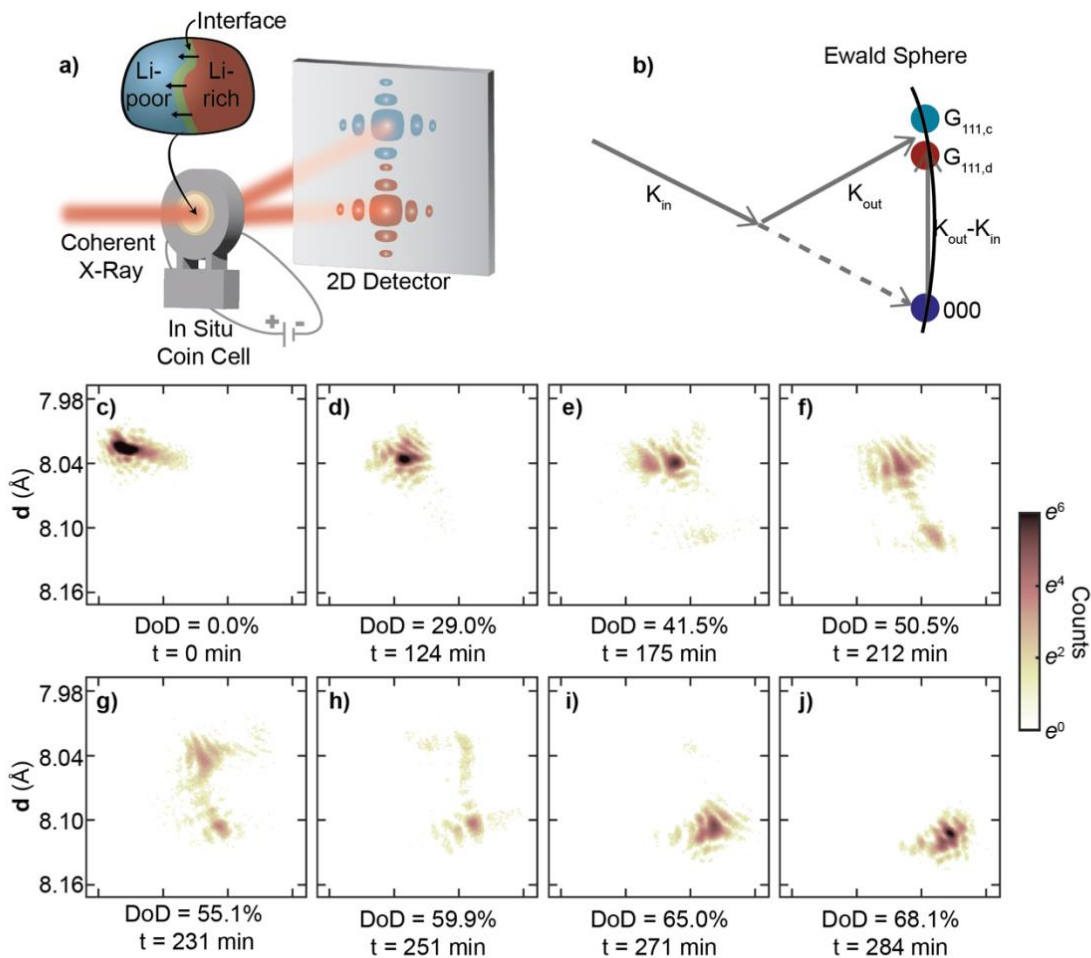

Figure 1. Operando Bragg Coherent Diffraction of a single $Li_xNi_{0.5}Mn_{1.5}O_4$ nanoparticle during the discontinuous solid-solid phase transformation induced by electrochemical Li-insertion. a) Experimental setup showing the operando coin cell, illuminated by coherent X-rays at 9 keV with a focus size of 800 nm, and diffraction around the 111 Bragg peak recorded on an area detector. During discharge, the $Li_xNi_{0.5}Mn_{1.5}O_4$ particle undergoes a discontinuous phase transformation via phase coexistence of the Li-poor phase (in blue) and Li-rich phase (in red), separated by the interface (in green). b) Schematic illustrating the Ewald sphere construction mapping out the interference profile surrounding the reciprocal lattice points $G_{111,d}$ (Li-rich phase) and $G_{111,c}$ (Li-poor phase). The area detector records a segment of the Ewald sphere. As it intersects both reciprocal space vectors, the detector image shows a split peak. c) – j) Cross-sections of the 3D diffraction pattern for the same $Li_xNi_{0.5}Mn_{1.5}O_4$ nanoparticle at various depths of discharge. The vertical axis shows the lattice spacing, $d$, determined as $2\pi/q$, where $q$ is the coordinate in reciprocal space.

Figure 1a shows the experimental setup for the operando measurement (see Methods and Figs.S1-2). During discharge, Li-ion intercalation induces a structural phase transformation from the Li-poor phase (smaller lattice constant, $d_c$) to the Li-rich phase (larger lattice constant, $d_d$)[13,24]. The different lattice spacing results in two separate Bragg reflections, $\mathbf{G}_{111,c}$ and $\mathbf{G}_{111,d}$. When the two phases coexist inside a single crystal, both Bragg reflections are present simultaneously (Fig. 1b). Due to the illumination from spatially coherent x-rays on a single particle, the scattering amplitude is decorated by an interference pattern[18]. When the coherent x-rays illuminate the particle with coexisting phases, the superposition of the interference patterns captures the spatial distribution and the relative crystallographic registry of both phases, as well as the structure of the interface. To measure the 3D interference patterns, we recorded a series of 2D sections of the Ewald sphere across the reciprocal space by rocking the *operando* cell (Fig.S3)[25].

Figure 1c-j presents the central slice of the measured *operando* 3D diffraction data as a function of depth of discharge (DoD) for the 111 Bragg peak of both phases converted to lattice spacing $d$ (see Fig. S4 for the full dataset with 50 measurements). At 0% DoD, corresponding to the fully charged state, only one Bragg peak surrounded by interference fringes is visible at larger $d$ (Fig.1c). During the initial stages of discharge, spanning from 0% to 35% DoD, the peak position is static, yet variations in the interference fringes suggest local structural changes within the nanoparticle (Fig.1d). Starting at 41.5% DoD (Fig.1e), a secondary peak emerges at larger $d$, steadily gaining intensity at the cost of diminishing intensity in the primary peak (Fig.1e-i). This is the hallmark of a discontinuous structural phase transformation, characterized by a substantial variation in the order parameter (namely, the lattice constant). The presence of both diffraction peaks emanating from an individual sub-micrometer crystal indicates phase coexistence. Notably, the presence of the coexisting peaks in the diffraction data occurs at the voltage plateau in the electrochemical data[26] (Fig.S5), consistent with previous *operando* x-ray diffraction[13,20].

To interpret the measurements, we use the recently developed correlated phase retrieval algorithm[23] to invert the diffraction data into real-space 3D images. Critical for the success of the algorithm is inverting a series of *operando* measurements simultaneously, while assuming an approximately static shape of the particle across different charge states (Fig.S6). We argue that the assumption is true here because the material is stable for hundreds of charge-discharge cycles, suggesting retention of the particle shape during a single discharge. The algorithm reconstructs the particle shape and the displacement field of the nanoparticle along the scattering vector (here, [111])[20,21,27]. Subsequently, we derive the 3D strain distribution from this displacement field through numerical differentiation along the scattering vector[28]. The reconstructed diffraction pattern exhibits good agreement with the measured diffraction data (Fig.S7), affirming the success of the phase retrieval process.

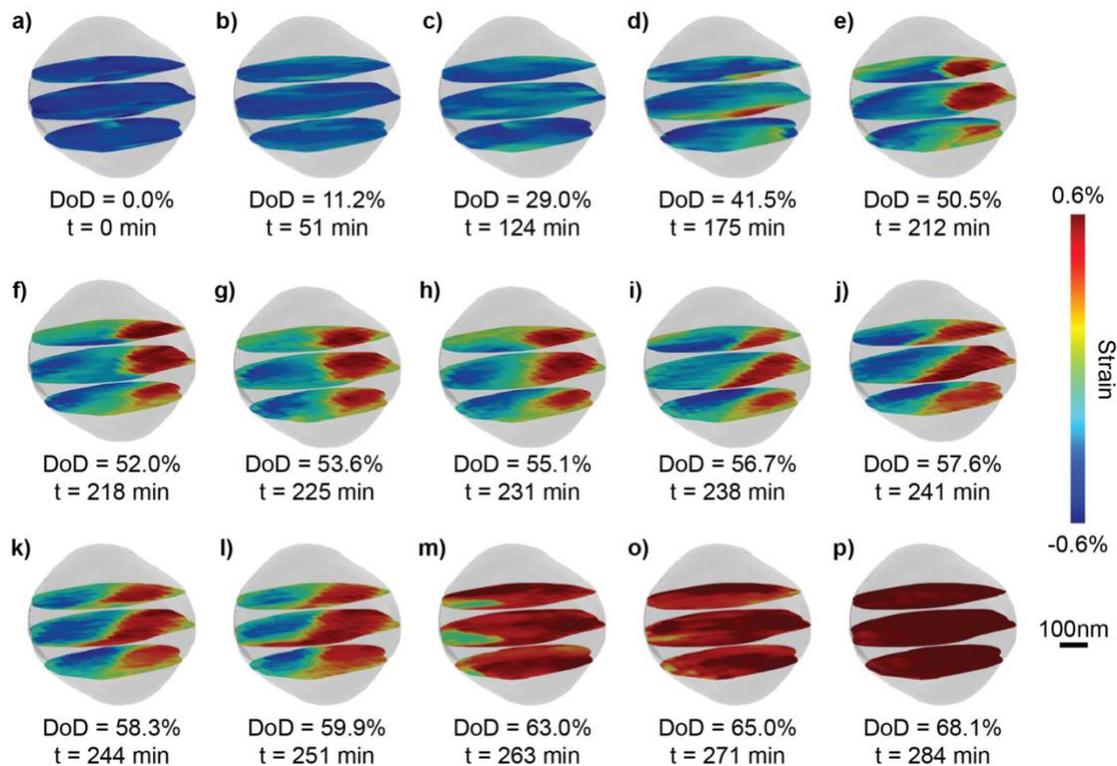

**Figure 2. Operando visualization of the coexisting phases during the structural phase transformation under lithium intercalation. a) – p)** The 3D strain field inside a single $Li_xNi_{0.5}Mn_{1.5}O_4$ nanoparticle during the two-phase reaction induced by Li intercalation. The strain maps are extracted by inverting coherent x-ray diffraction data shown in Figure 1 c-j and Fig.S4. The semi-transparent isosurface portrays the particle shape, while the colored slices display the strain distribution, $\varepsilon_{111}(r)$, illustrated on three chosen planes. We show the strain field, $\varepsilon_{111}(r)$, as the local lattice constant compared to the average lattice constant, $d_a$, of the (111) planes between the fully charged, Li-poor ($d_c$) and the partially discharged, Li-rich ($d_d$) phase, where $d_a = (d_c + d_c)/2$ and $\varepsilon_{111}(r) = d(r)/d_a - 1$. At 0% DoD in **a)**, the particle has a uniform negative strain shown in blue, corresponding to the Li-poor phase with a small lattice constant, $d_c$. At the end of the two-phase reaction around 68.1% DoD in **p)**, the particle presents a uniform positive strain shown in red, corresponding to the Li-rich phase with a large lattice constant $d_d$.

Figure 2 illustrates the *operando* strain evolution within a $Li_xNi_{0.5}Mn_{1.5}O_4$ nanoparticle, as obtained from the phase retrieval on the *operando* diffraction data (see Fig.S8 for the full dataset of 50 images). Overall, the imaging data portrays the nucleation and growth of the Li-rich phase at the expense of the Li-poor phase through interface advancement (similar behavior is observed in another $Li_xNi_{0.5}Mn_{1.5}O_4$ particle, see Figs.S11-12). At the onset of discharge, the nanoparticle consists of an almost homogeneous Li-poor phase (negative strain, depicted in blue) (Fig.2a-c). As the electrochemical lithiation proceeds, an inclusion of the Li-rich phase (positive strain, depicted in red) nucleates at the bottom right corner of the nanoparticle (Fig.2d). At this stage, the imaging suggests the presence of multiple nucleation sites (Fig.S9); however, as the growth proceeds, only a single inclusion becomes visible (Fig.2e). The merging is likely driven by surface tension, akin to Ostwald ripening or coarsening[29]. Throughout the subsequent lithiation, the Li-rich phase grows at the expense of the Li-poor phase via interface propagation (Fig.2e-o). We approximate the interface propagation velocity to be around 0.13 nm/s, notably slower than the mean velocity of Li-ions due to diffusion (Fig.S10). Consequently, the Li ions have sufficient time to equilibrate concentration gradients within each phase, likely resulting in a sharp Li-concentration gradient at the interface[30]. This is consistent with the interface width between the coexisting structures in Fig.2i-j, which is less than 100 nm, approaching the spatial resolution of *operando* BCDI[21]. By the end of the process, the nanoparticle comprises entirely of the Li-rich phase (Fig.2p).

In addition to visualizing in real time the phase distribution and the interface propagation during the phase transformation, *operando* BCDI provides insights into the morphological evolution of the interface. Our data shows an initially curved interface (Fig.2e-h) that subsequently transforms into a planar configuration (Fig.2j-k). These morphological changes are likely associated with the microstructure of the interface between the Li-poor and the Li-rich phases at different DoDs. Heterointerfaces between two distinct crystalline structural phases are typically classified into three types: coherent interfaces maintaining complete continuity of the lattice (Fig.3a), semi-coherent interfaces with piecewise continuity separated by misfit dislocations (Fig.3d), and incoherent interfaces with no registry between the two phases[31]. In our single particle diffraction data, the coexisting phases generate diffraction peaks in proximity (Fig.1e-h). Therefore, the two crystalline phases are closely aligned, ruling out a fully incoherent interface, which typically occurs when the crystal planes are misaligned by more than 15 degrees[32].

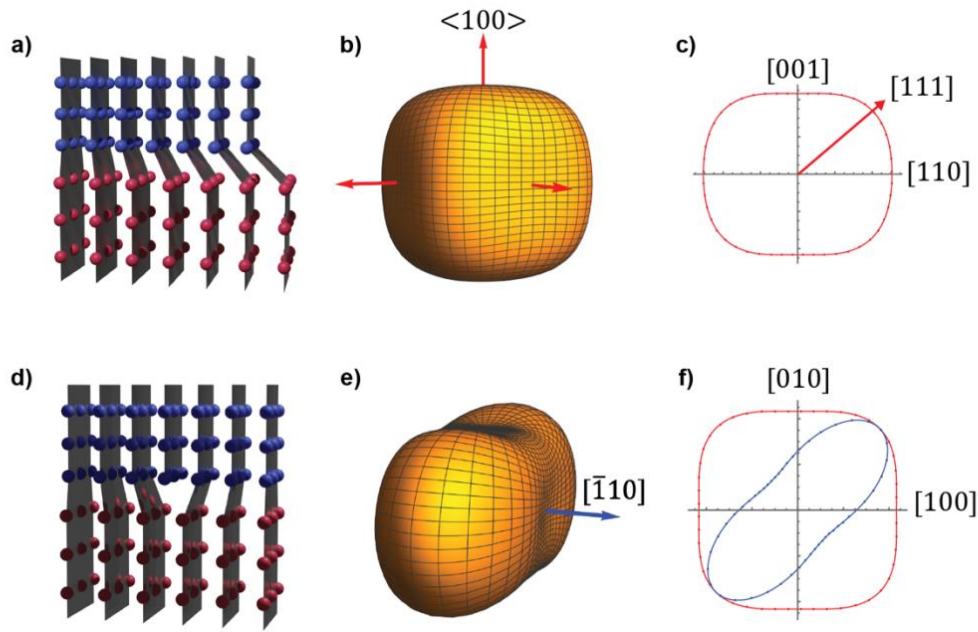

**Figure 3. Microelasticity theory for coherent and semi-coherent interface in $Li_xNi_{0.5}Mn_{1.5}O_4$.** **a)** Illustration showing a coherent interface. Each plane of blue atoms representing the Li-poor phase is connected to a plane of the red atoms representing the Li-rich phase. The lattice distortion gradually intensifies from the left to the right. **b)** Isosurface plot of the elastic strain energy as a function of normal direction for a coherent interface. The red arrows indicate the energy minima directions, <100>. **c)** 2D parametric plot of the strain energy illustrated in **b)** in the plane spanning the [001] and [110] directions. The energy minimum, [001], and maximum direction, [111], display similar values. **d)** Illustration of a semi-coherent interface with a misfit edge dislocation – one extra half-plane inserted from the top – that relieves the misfit strain in the direction of the Burgers vector perpendicular to the half-plane. **e)** Isosurface plot of elastic strain energy for a semi-coherent interface with coherency loss in the [110] direction. The blue arrow indicates the energy minimum direction along [$\bar{1}$10]. **f)** 2D parametric plot in the plane spanning [100] and [010], comparing strain energy between a coherent (red) and a semi-coherent (blue) interface. The semi-coherent interface exhibits significant anisotropy in its strain energy.

To distinguish between the coherent and semi-coherent heterointerface, we adopt the microelasticity model within the framework of continuum mechanics[33], which was adapted to phase transformation in $Li_xFePO_4$ (0<x<1), another technically important phase separating cathode material[34,35]. Specifically, we approximate the elastic strain energy, denoted as **B**, for an inclusion of the Li-rich phase inside the Li-poor phase (using Einstein notation for summation)

$$\boldsymbol{B}(\vec{n}) = \lambda_{ijkl}\varepsilon_{ij}^0\varepsilon_{kl}^0 - \vec{n_i}\sigma_{ij}^0\Omega_{jl}(\vec{n})\sigma_{lm}^0\vec{n_m} \qquad (1)$$

where $\vec{n}$ is the interface normal, $\varepsilon_{ij}^0$ is the strain tensor, $\sigma_{ij}^0$ is the stress tensor, $\Omega_{ij}$ is related to the elastic Green's tensor and defined as $\Omega_{ij}^{-1} = \lambda_{iklj}\vec{n}_k\vec{n}_l$, and $\lambda_{ijkl}$ is the elastic stiffness tensor (Fig. S13). We first consider a coherent interface separating the two phases with an isotropic transformation strain of 1%. The calculated elastic strain energy from equation (1) reveals an energy minimum direction along the family of <100> directions (Fig.3b-c). Notably, the strain energy along the maximum direction, <111>, differs from the minimum direction, <100>, by a mere 8%. Given this weak dependence of the elastic energy on the interface orientation, the modeling is consistent with an overall spherical shape of the interface, as shown in Fig.2e-h.

As the structural transformation advances and the interfacial area expands, significant coherency strain arises due to the lattice mismatch between the two phases. To alleviate this strain, the interface is expected to introduce misfit dislocations and form a semi-coherent interface (Fig.3d). In this configuration, the interface comprises regions that maintain lattice continuity but is interspersed with dislocation cores that release interfacial strain and disrupt this continuity[31]. These misfit dislocations introduce an anisotropy to the otherwise isotropic transformation strain, consequently altering the orientations of the low-energy phase boundaries[34,35]. The transformation from coherent to semi-coherent interface is reminiscent of the coherency loss in metals[36]. For the spinel structure, the slip directions and planes are $<110>$ and $\{111\}$[37]. We assume that the dislocations possess a Burgers vector along [110] and lie within the $(\bar{1}11)$ plane (Fig. S13). As a result, these dislocations lead to a loss of coherency along [110] by providing strain relief in this direction. Setting $\varepsilon_{110} = 0$ results in the minimum elastic strain energy, $\boldsymbol{B}(\vec{n})$, along $[\bar{1}10]$ (Fig.3e), and the energy difference between the lowest and highest energy directions has increased to 197% (Fig.3f). This anisotropy results in a preferential interface orientation along $[\bar{1}10]$, and correspondingly, the microelasticity model predicts a transition of the interface geometry from a curved to a planar configuration, as we observe in Fig.2h-i.

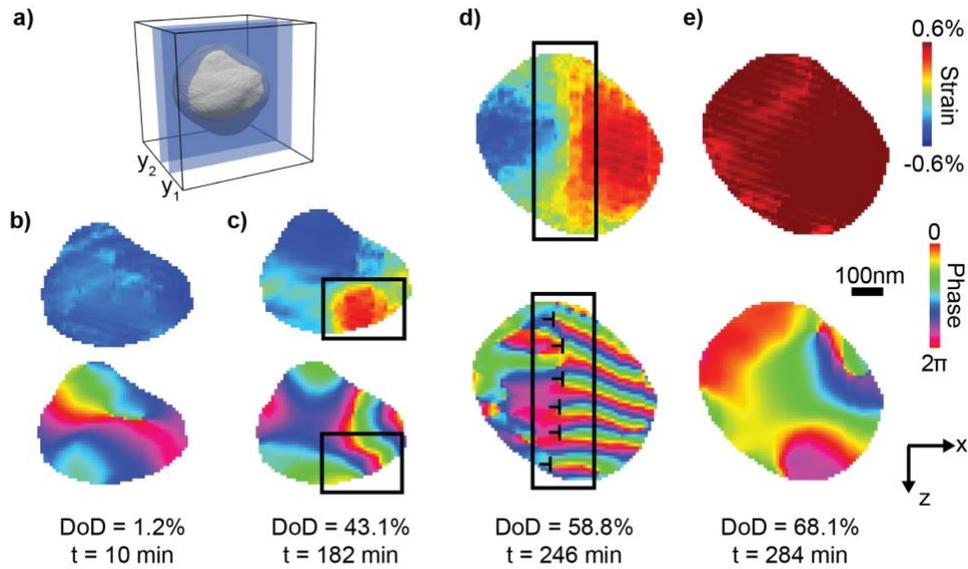

**Figure 4. The cross-sections of the reconstructed strain (top) and displacement (bottom) field in the *xz* plane at different depths of discharge. a)** Three-dimensional representation of the LNMO particle aligned at the same angle as in Fig.2, with the two slices shown as strain and displacement field in **b)** – **e)**. **b)** Prior to the phase transformation, the strain map corresponds to a fully charged Li-poor phase. The false color in the phase maps reflects the displacement from the ideal lattice at each position. **c)** At the early stage of the phase transformation, the Li-rich phase (enclosed by the rectangle) nucleates at the lower right corner, leading to an intensified color gradient within the rectangle. Both *xz* slices in **b)** and **c)** are located at $y_1$ in **a)**. **d)** During the phase transformation, the nucleated Li-rich phase forms a semi-coherent interface with the Li-poor phase (outlined by the rectangle), which contains an array of dislocations each denoted by ⊥ in the phase map. The dislocations run into the page and propagate along the negative x-direction with the interface. **e)** At the end of phase transformation, the strain map shows the Li-rich phase with no observable signs of dislocations in the phase map. The *xz* slices in **d)** and **e)** are located at $y_2$ in **a)**. The scattering vector, **q**, points along the *z*-axis.

To validate the transition from a coherent to a semi-coherent interface, we examine if an array of misfit dislocations is visible at the interface after it transforms into a planar geometry. Dislocations can be directly observed in the displacement field from BCDI data as singularities[20,21,27]. At the fully charged state (DoD = 1.2%), the particle exhibits uniform negative strain, and the displacement field is continuous without singularities (Fig.4a). As the discharge progresses, the Li-rich phase with positive strain forms, leading to changes in the displacement field (Fig.3e). At the interface, the continuity of the displacement field remains uninterrupted, showing no indications of structural defects such as dislocations. Yet, for the subsequent discharge, when the planar interface forms, the displacement map in Fig.4c reveals an array of misfit dislocations. In this diffraction geometry, we measure the projection of the Burgers vector along [111]. Thus, a phase discontinuity of $2\pi$ from a dislocation with Burgers vector along [110] is expressed as a phase discontinuity of 5.1 radians along [111] (Fig.S14). Given a 1% lattice mismatch, we anticipate observing one dislocation per 100 unit cells, which translates to approximately one dislocation per 65 nm, on the same order of magnitude as the observed dislocation density of about one per 95 nm in Fig.4c. The presence of these dislocations aligns with our earlier hypothesis regarding the introduction of a semi-coherent interface for strain relaxation in the microelasticity model. The preferential [1̄10] energy minimum direction is also evidenced in Fig.4c as the strain map shows the interface orientation almost normal to the scattering vector [111].

**Discussion**

Various models have been proposed to explain intercalation-driven phase transformations in rechargeable battery electrodes. The 'shrinking core' model applied to $Li_xFePO_4$ assumes uniform Li insertion in isotropic spherical particles[38–40]. Built on the anisotropic Li-ion diffusion in $Li_xFePO_4$, the intercalation wave mechanism (or "domino-cascade") considers an interfacial zone sweeping through the particle[41,42]. Incorporating both models, Singh et al. develops a general continuum theory and categorizes intercalation dynamics based on crystal anisotropy and ion insertion kinetics across electrode/electrolyte interface (namely isotropic bulk transport limited, anisotropic bulk transport limited, isotropic surface reaction limited, and anisotropic surface reaction limited)[30]. In $Li_xNi_{0.5}Mn_{1.5}O_4$, the lithium diffusion is isotropic[43]. Thus, at (dis)charge rates we measured here (C/10), the phase transition falls under isotropic surface reaction limited regime. Our experiments suggest a nucleation and growth mechanism for the transformation, consistent with findings by Kuppan et al. using spectroscopic ex-situ imaging in chemically oxidized ordered $Li_xNi_{0.5}Mn_{1.5}O_4$ single crystals[44]. Within this framework, the nucleus grows from a single localized point within the particle and extends through interface propagation.

In addition to evaluating different models for phase growth, access to crystal structure in our *operando* data provides insights into the nanomechanics at the propagating interface. The direction of dislocation motion has a significant component perpendicular to the Burgers vector. This motion is reminiscent of dislocation climb, a process requiring diffusion of host species (Ni, Mn, and O). Yet, given the reversibility of hundreds of cycles in $Li_xNi_{0.5}Mn_{1.5}O_4$[45], a diffusional flux of host species during every cycle seems unlikely. Drawing inspiration from the well-established theory of phase transformations in metals[31], we propose a model involving a glissile motion of interfacial dislocations. In this framework, the interface is generated through an invariant shear deformation to accommodate the lattice mismatch between the two phases. As the interface propagates, dislocations enter the crystal at the surface along the slip planes created by the shear (Fig.S15)[32], maintaining a conservative glissile interface motion with no host species diffusion, which allows fast kinetics[31]. *Operando* imaging of a $Li_xNi_{0.5}Mn_{1.5}O_4$ nanoparticle measured at C/2 discharge rate shows that the two-phase reaction completes within 40 minutes (Fig.S16), five times faster than the transformation at C/10. Despite the rapid reaction, phase separation is still evident in the strain maps (Fig.S17), revealing that the two-phase transformation kinetics can be fast. Our results are consistent with recent phase-field modeling that suggests interfacial coherency loss can substantially improve reaction kinetics for high-rate cathode materials when the two-phase reaction is unavoidable[35]. When the phase transformation completes, the interfacial dislocations move along with the interface and exit the nanoparticle (Fig.4d).

In conclusion, we have successfully imaged a discontinuous phase transformation inside a single battery electrode nanoparticle in a fully operational device. The 3D snapshots of the crystal structure with a spatial resolution in the tens of nanometers and a time resolution in a few minutes reveal a nucleation and growth transformation mechanism. At C/10 and C/2 discharge rate, we see no kinetic limitations due to loss of interface coherency from misfit dislocations. Although our current time resolution with BCDI limits the study for faster charge rates, ongoing developments of diffraction-limited synchrotron sources will boost BCDI to seconds[46]. The advancement will enable imaging of rapid, discontinuous phase transformations and investigating effects like interface pinning that we observed in Fig.2l. *Operando* observations of interface dynamics warrant further exploration into the atomic arrangement of interface structures. While BCDI measures the long-range displacement field at the interface, a multimodal approach, combining *operando* BCDI with atomistic simulations and high-resolution electron microscopy, is imperative. Eventually, imaging phase transformations in nanoparticulate functional materials embedded in multicomponent devices will establish a feedback loop between stimuli and characterization. This approach is critical for optimizing properties like durability and ionic transport in materials exhibiting discontinuous phase transformations.

## Methods
### Synthesis and coin cell assembly
$Li_\delta Ni_{0.5}Mn_{1.5}O_{4-\delta}$ disordered spinel was synthesized using the sol-gel method[39]. The coin cell was assembled using LNMO as the cathode and graphite as the anode (Fig.S2). A 3 mm opening was created around the center of the base and top shells and sealed with a Kapton film to allow for X-ray transmission.

### Bragg Coherent Diffractive Imaging experiment
The operando BCDI experiment was conducted at Sector 34-ID-C in the Advanced Photon Source at Argonne National Laboratory. A double crystal monochromator was used to select x-rays with energy $E = 9$ keV. The coherent X-rays with a focus size of 800 nm were incident on a fully operational half-cell. The rocking curve around the 111 Bragg peak was collected by a 2D detector (Timepix, 256 × 256 pixels, each pixel 55µm × 55µm) around $2\theta = 17$ degrees ($\Delta\theta = \pm 0.3°$). The detector was placed 1.1 meter away from the sample and an evacuated flight tube was inserted between the sample and the detector. A total of 76 diffraction patterns were collected for a single 3D rocking scan with 1 second exposure time for each image. The 3D diffraction pattern of the same particle was continuously captured under *operando* conditions, while the coin cell battery was discharging at C/10 or C/2. The low discharge rate was chosen to ensure that the particle remained in its discharge state throughout the scan. Each scan had a duration of about 2 minutes. Two rounds of alignment scans in the labx, labz (sample position relative to the incident beam) and theta directions were taken between every three 3D rocking scans to ensure that the particle did not move out of the beam or rotate away from the diffraction condition.

### Inversion of x-ray data using phase retrieval
The details of the correlated data inversion algorithm were reported elsewhere[23]. Here, for brevity we summarize the main aspects of the algorithm we used to invert the *operando* data. 10 diffraction scans that describe the entire two-phase reaction were inverted simultaneously. Every diffraction scan was aligned such that the center of each diffraction had the same scattering vector (during experiment, the scattering angle was shifted to follow the peak evolution). The data inversion started with each scan running 30 reconstructions individually, each initiated with a random phase. For every 10 iterations, the error matrix of the correlation within the 30 reconstructions was calculated for each scan. Then we averaged the support of the 5 best correlated reconstructions in each scan across all 10 scans (in total 50 reconstructions) while leaving the displacement fields unchanged. Then the averaged support was multiplied by the individual support and became the input support for the next set of the reconstructions. The reconstruction consisted of a total of 610 iterations with alternating 10 iterations of the ER algorithm and followed by 50 iterations of the RAAR algorithm. For the primary particle in the main text, five datasets that contain mutually different scans were reconstructed and then stitched back together into a single sequence. The series shown in Fig. 2 and Fig. S8 is a collage of multiple independent reconstruction procedures, i.e, some have no overlap of the diffraction data taken for the reconstructions. We take the continuous evolution of the nucleation and growth as a testament for the robustness of the algorithm. For the supplementary particle, one dataset that consists of 13 scans and one dataset that consists of 15 scans were reconstructed. The algorithm was performed 10 times on each dataset and the final imaging is the result of 5 × 10 reconstructions of each scan.

**Acknowledgements**
The work at Cornell was supported by the National Science Foundation under award number CAREER DMR-1944907. The work at UC San Diego was supported by the Sustainable Power and Energy Center (SPEC) and US Department of Energy, Office of Science, Office of Basic Energy Sciences, under contract No. DE-SC0001805. This research used resources of the Advanced Photon Source, a U.S. Department of Energy (DOE) Office of Science user facility operated for the DOE Office of Science by Argonne National Laboratory under Contract No. DE-AC02-06CH11357.

**Reference**

1. Gibbs, J. W. *On the Equilibrium of Heterogeneous Substances*. vol. 2 300–320 https://archiv.ub.uni-heidelberg.de/volltextserver/13220/ (1879).

2. Darul, J., Nowicki, W., Piszora, P., Baehtz, C. & Wolska, E. Synchrotron X-ray powder diffraction studies on the order–disorder phase transition in lithium ferrites. *J. Alloys Compd.* **401**, 60–63 (2005).

3. Krauss, G. Martensite in steel: strength and structure. *Mater. Sci. Eng. A* **273–275**, 40–57 (1999).

4. Easterling, D. A. P., Kenneth E. Easterling, Kenneth E. *Phase Transformations in Metals and Alloys (Revised Reprint)*. (CRC Press, 2009). doi:10.1201/9781439883570.

5. Wilde, G. Structural Phase Transformations in Nanoscale Systems. *Adv. Eng. Mater.* **23**, 2001387 (2021).

6. Tang, M., Carter, W. C. & Chiang, Y.-M. Electrochemically Driven Phase Transitions in Insertion Electrodes for Lithium-Ion Batteries: Examples in Lithium Metal Phosphate Olivines. *Annu. Rev. Mater. Res.* **40**, 501–529 (2010).

7. Mariyappan, S., Wang, Q. & Tarascon, J. M. Will Sodium Layered Oxides Ever Be Competitive for Sodium Ion Battery Applications? *J. Electrochem. Soc.* **165**, A3714 (2018).

8. Sharma, N. *et al.* Direct Evidence of Concurrent Solid-Solution and Two-Phase Reactions and the Nonequilibrium Structural Evolution of LiFePO 4. (2012) doi:10.1021/ja301187u.

9. Wagemaker, M. *et al.* A Kinetic Two-Phase and Equilibrium Solid Solution in Spinel Li4+xTi5O12. *Adv. Mater.* **18**, 3169–3173 (2006).

10. Ganapathy, S., Vasileiadis, A., Heringa, J. R. & Wagemaker, M. The Fine Line between a Two-Phase and Solid-Solution Phase Transformation and Highly Mobile Phase Interfaces in Spinel Li4+xTi5O12. *Adv. Energy Mater.* **7**, (2017).

11. Liang, G., Peterson, V. K., See, K. W., Guo, Z. & Kong Pang, W. Developing high-voltage spinel LiNi 0.5 Mn 1.5 O 4 cathodes for high-energy-density lithium-ion batteries: current achievements and future prospects. *J. Mater. Chem. A* **8**, 15373–15398 (2020).

12. Zhao, H. *et al.* Cobalt-Free Cathode Materials: Families and their Prospects. *Adv. Energy Mater.* **12**, 2103894 (2022).

13. Singer, A. *et al.* Nonequilibrium structural dynamics of nanoparticles in LiNi1/2Mn3/2O4 cathode under operando conditions. *Nano Lett.* **14**, 5295–5300 (2014).

14. Huang, J. J. *et al.* Disorder Dynamics in Battery Nanoparticles During Phase Transitions Revealed by Operando Single-Particle Diffraction. *Adv. Energy Mater.* **12**, 2103521 (2022).

15. Liu, H. *et al.* Capturing metastable structures during high-rate cycling of LiFePO4 nanoparticle electrodes. *Science* **344**, 1252817 (2014).

16. Lin, F. *et al.* Synchrotron X-ray Analytical Techniques for Studying Materials Electrochemistry in Rechargeable Batteries. *Chem. Rev.* **117**, 13123–13186 (2017).

17. Pfeifer, M. A., Williams, G. J., Vartanyants, I. A., Harder, R. & Robinson, I. K. Three-dimensional mapping of a deformation field inside a nanocrystal. *Nature* **442**, 63–66 (2006).

18. Robinson, I. & Harder, R. Coherent X-ray diffraction imaging of strain at the nanoscale. *Nat. Mater.* **8**, 291–298 (2009).

19. Clark, J. N. *et al.* Three-dimensional imaging of dislocation propagation during crystal growth and dissolution. *Nat. Mater.* **14**, 780–784 (2015).

20. Ulvestad, U. *et al.* Topological defect dynamics in operando battery nanoparticles. *Science* **348**, 1344–1347 (2015).

21. Singer, A. *et al.* Nucleation of dislocations and their dynamics in layered oxide cathode materials during battery charging. *Nat. Energy* **3**, 641–647 (2018).

22. Diao, J. *et al.* Evolution of ferroelastic domain walls during phase transitions in barium titanate nanoparticles. *Phys. Rev. Mater.* **4**, 106001 (2020).

23. Wang, Z., Gorobtsov, O. & Singer, A. An algorithm for Bragg coherent x-ray diffractive imaging of highly strained nanocrystals. *New J. Phys.* **22**, (2020).

24. Kim, J.-H., Yoon, C. S., Myung, S.-T., Prakash, J. & Sun, Y.-K. Phase Transitions in Li$_{1-\delta}$Ni$_{0.5}$Mn$_{1.5}$O$_4$ during Cycling at 5 V. *Electrochem. Solid-State Lett.* **7**, A216 (2004).

25. Williams, G. J., Pfeifer, M. A., Vartanyants, I. A. & Robinson, I. K. Three-Dimensional Imaging of Microstructure in Au Nanocrystals. *Phys. Rev. Lett.* **90**, 4 (2003).

26. Van der Ven, A., Bhattacharya, J. & Belak, A. A. Understanding Li Diffusion in Li-Intercalation Compounds. *Acc. Chem. Res.* **46**, 1216–1225 (2013).

27. Sun, Y. *et al.* X-ray Nanoimaging of Crystal Defects in Single Grains of Solid-State Electrolyte Li$_{7-3x}$Al$_x$La$_3$Zr$_2$O$_{12}$. *Nano Lett.* **21**, 4570–4576 (2021).

28. Ulvestad, A. *et al.* Single particle nanomechanics in operando batteries via lensless strain mapping. *Nano Lett.* **14**, 5123–5127 (2014).

29. Sethna, J. P. *Entropy, Order Parameters, and Complexity*. (Oxford University Press, 2022).

30. Singh, G. K., Ceder, G. & Bazant, M. Z. Intercalation dynamics in rechargeable battery materials: General theory and phase-transformation waves in LiFePO4. *Electrochimica Acta* **53**, 7599–7613 (2008).

31. Sutton, A. P. & Balluffi, R. W. *Interfaces in Crystalline Materials*. (Oxford University Press, 2007).

32. Balluffi, R. W., Allen, S. M. & Carter, W. C. *Kinetics of Materials*. (John Wiley & Sons, 2005).

33. Khachaturyan, A. G. *Theory of structural transformations in solids*. (Courier Corporation, 2013).

34. Cogswell, D. A. & Bazant, M. Z. Coherency strain and the kinetics of phase separation in LiFePO 4 nanoparticles. *ACS Nano* **6**, 2215–2225 (2012).

35. Heo, T. W., Tang, M., Chen, L.-Q. & Wood, B. C. Defects, Entropy, and the Stabilization of Alternative Phase Boundary Orientations in Battery Electrode Particles. *Adv. Energy Mater.* **6**, 1501759 (2016).

36. Shi, R., Ma, N. & Wang, Y. Predicting equilibrium shape of precipitates as function of coherency state. *Acta Mater.* **60**, 4172–4184 (2012).

37. Callister, W. D. *Fundamentals of materials science and engineering: an integrated approach*. (John Wiley & Sons, cop., 2013).

38. Padhi, A. K., Nanjundaswamy, K. S. & Goodenough, J. B. Phospho-olivines as Positive-Electrode Materials for Rechargeable Lithium Batteries. *J. Electrochem. Soc.* **144**, 1188 (1997).

39. Andersson, A. S., Kalska, B., Häggström, L. & Thomas, J. O. Lithium extraction/insertion in LiFePO4: an X-ray diffraction and Mössbauer spectroscopy study. *Solid State Ion.* **130**, 41–52 (2000).

40. Srinivasan, V. & Newman, J. Discharge Model for the Lithium Iron-Phosphate Electrode. *J. Electrochem. Soc.* **151**, A1517 (2004).

41. Delmas, C., Maccario, M., Croguennec, L., Le Cras, F. & Weill, F. Lithium deintercalation in LiFePO4 nanoparticles via a domino-cascade model. *Nat. Mater. 2008 78* **7**, 665–671 (2008).

42. Bazant, M. Z. Theory of Chemical Kinetics and Charge Transfer based on Nonequilibrium Thermodynamics. *Acc. Chem. Res.* **46**, 1144–1160 (2013).

43. Xu, B. & Meng, S. Factors affecting Li mobility in spinel LiMn2O4—A first-principles study by GGA and GGA+U methods. *J. Power Sources* **195**, 4971–4976 (2010).

44. Kuppan, S., Xu, Y., Liu, Y. & Chen, G. Phase transformation mechanism in lithium manganese nickel oxide revealed by single-crystal hard X-ray microscopy. *Nat. Commun.* **8**, (2017).

45. Cho, H.-M., Chen, M. V., MacRae, A. C. & Meng, Y. S. Effect of Surface Modification on Nano-Structured LiNi0.5Mn1.5O4 Spinel Materials. *ACS Appl. Mater. Interfaces* **7**, 16231–16239 (2015).

46. Raimondi, P. ESRF-EBS: The Extremely Brilliant Source Project. *Synchrotron Radiat. News* **29**, 8–15 (2016).

# Operando real-space imaging of a structural phase transformation in a high-voltage electrode: Supplementary Information

*Yifei Sun[1], Sunny Hy[2], Nelson Hua[3,4], James Wingert[3], Ross Harder[5], Ying Shirley Meng[2,6], Oleg Shpyrko[3], Andrej Singer[1,*]*

[1]*Department of Materials Science and Engineering, Cornell University, Ithaca, New York 14850, USA*
[2]*Department of Nanoengineering, University of California San Diego, La Jolla, California, 92093, USA*
[3]*Department of Physics, University of California San Diego, La Jolla, California, 92093, USA*
[4]*Laboratory for X-ray Nanoscience and Technologies, Paul Scherrer Institut, 5232 Villigen, Switzerland*
[5]*Advanced Photon Source, Argonne National Laboratory, Argonne, Illinois, 60439, USA*
[6]*Pritzker School of Molecular Engineering, University of Chicago, Chicago, Illinois, 60637, USA*
[*]asinger@cornell.edu

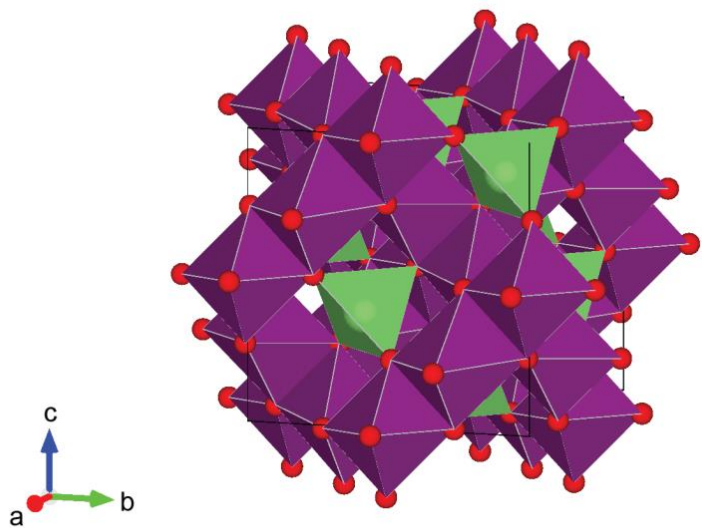

**Figure S1. Crystal structure of the disordered $Li_xNi_{0.5}Mn_{1.5}O_4$.** The disordered $Li_xNi_{0.5}Mn_{1.5}O_4$ has the cubic spinel structure with space group $Fd\bar{3}m$. The Ni and Mn reside within the purple octahedra with Ni occupying 25% of the sites and Mn occupying 75% of the sites. The Li resides within the green tetrahedra. The red atoms are oxygens.

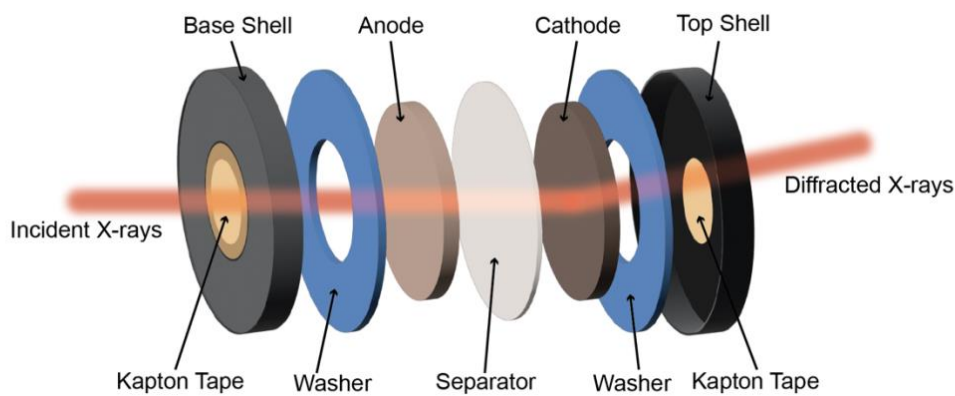

**Figure S2. Expanded view of the operando coin cell.** The cell top and base are from the standard CR2032 cells, which have a diameter of 20mm and a height of 3.2mm. Both sides have a hole drilled at the center of size around 3 mm in diameter, which is sealed by Kapton tape. The cell is placed so that the material of interest, the $Li_xNi_{0.5}Mn_{1.5}O_4$ cathode nanoparticles, is located downstream from the incident X-rays. The anode is lithium metal, and the separator (Celgard C480) contains electrolyte of 1 M solution of lithium hexafluorophosphate ($LiPF_6$) in a 1:1 volume mixture of ethylene carbonate (EC) and dimethyl carbonate (DMC).

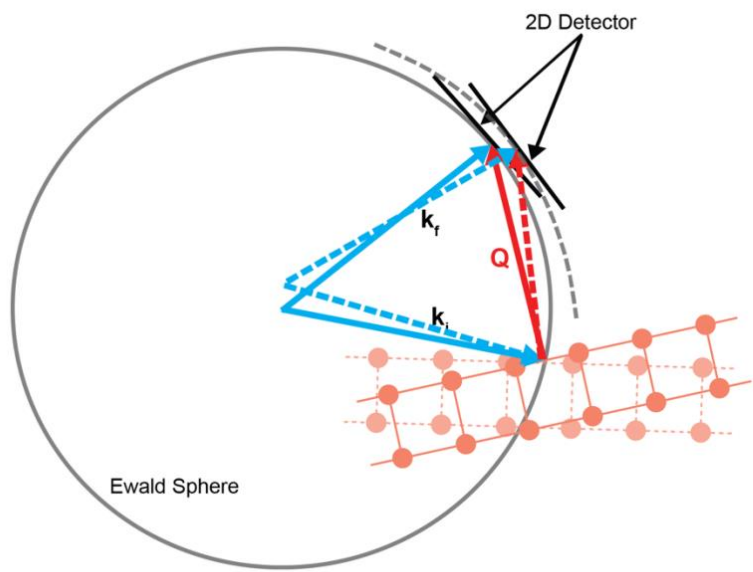

**Figure S3. Ewald sphere construction.** Rocking the crystal, shown as lattice plane in red, results in the Ewald sphere slicing the vicinity of the Bragg peaks at different positions in the reciprocal space. This is equivalent to shifting the detector perpendicular to the Ewald sphere around the Bragg peaks. The nearly parallel slices were combined to obtain a 3D pattern.[1]

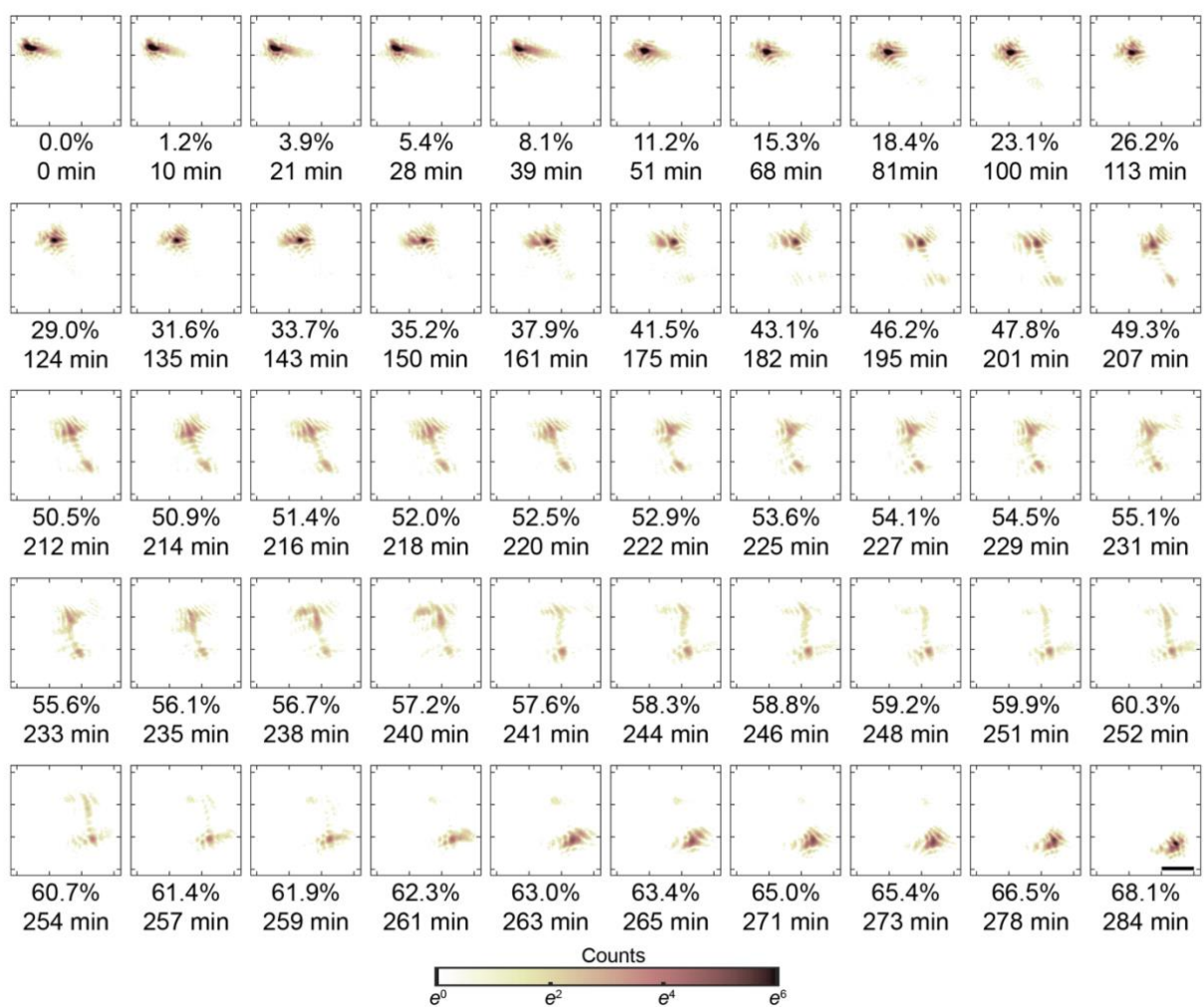

**Figure S4. Full set of the operando diffraction data from a single $Li_xNi_{0.5}Mn_{1.5}O_4$ particle undergoing a structural phase transformation during discharge.** Diffraction data shows the decrease of intensity of the lithium-poor phase (larger 2θ) and the increase of intensity of the lithium-rich phase (smaller 2θ). During the transition, two peaks coexist, indicating the presence of both phases inside the $Li_xNi_{0.5}Mn_{1.5}O_4$ nanoparticle. The two-phase reaction ends around when the battery is 68% discharged. The scalebar is 0.08 $nm^{-1}$. The diffraction data is inverted using the correlated phase retrieval algorithm.

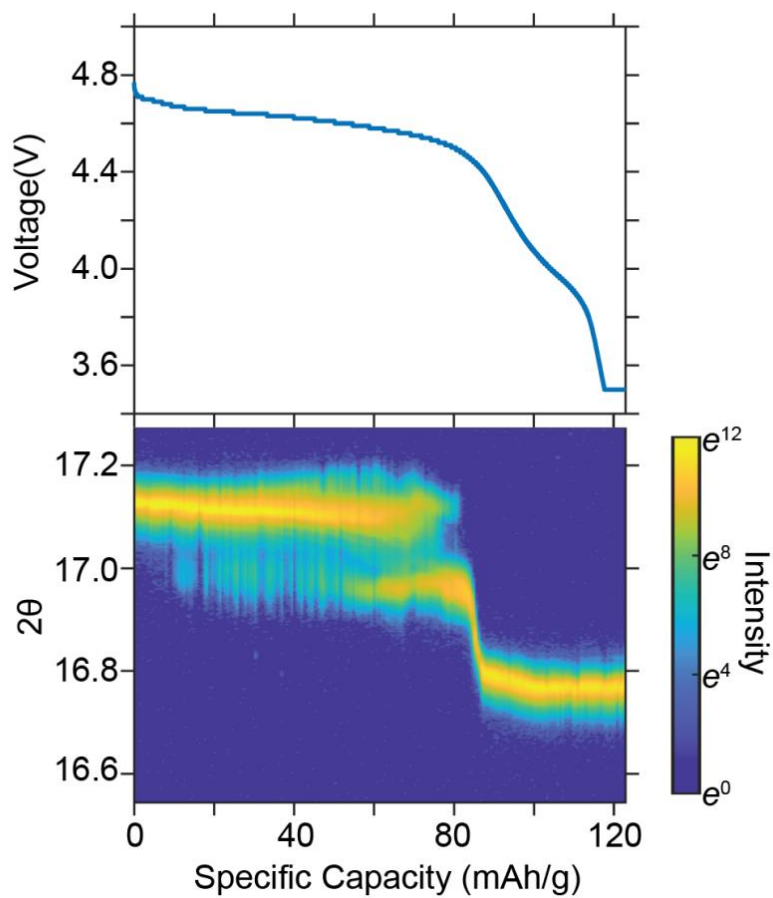

**Figure. S5. Electrochemical data of the cell (top) and diffraction data of a single $Li_xNi_{0.5}Mn_{1.5}O_4$ particle within the cell (bottom).** The voltage plateau in the electrochemical data coincides with the two-peak coexistence region in the diffraction data. This indicates that the electrochemistry behavior of the $Li_xNi_{0.5}Mn_{1.5}O_4$ particle that we selected for reconstruction is representative for all active material. When the specific capacity reaches around 80 mAh/g, the cell enters the solid-solution regime, which is reflected as one diffraction peak continuously shifting its 2θ position.

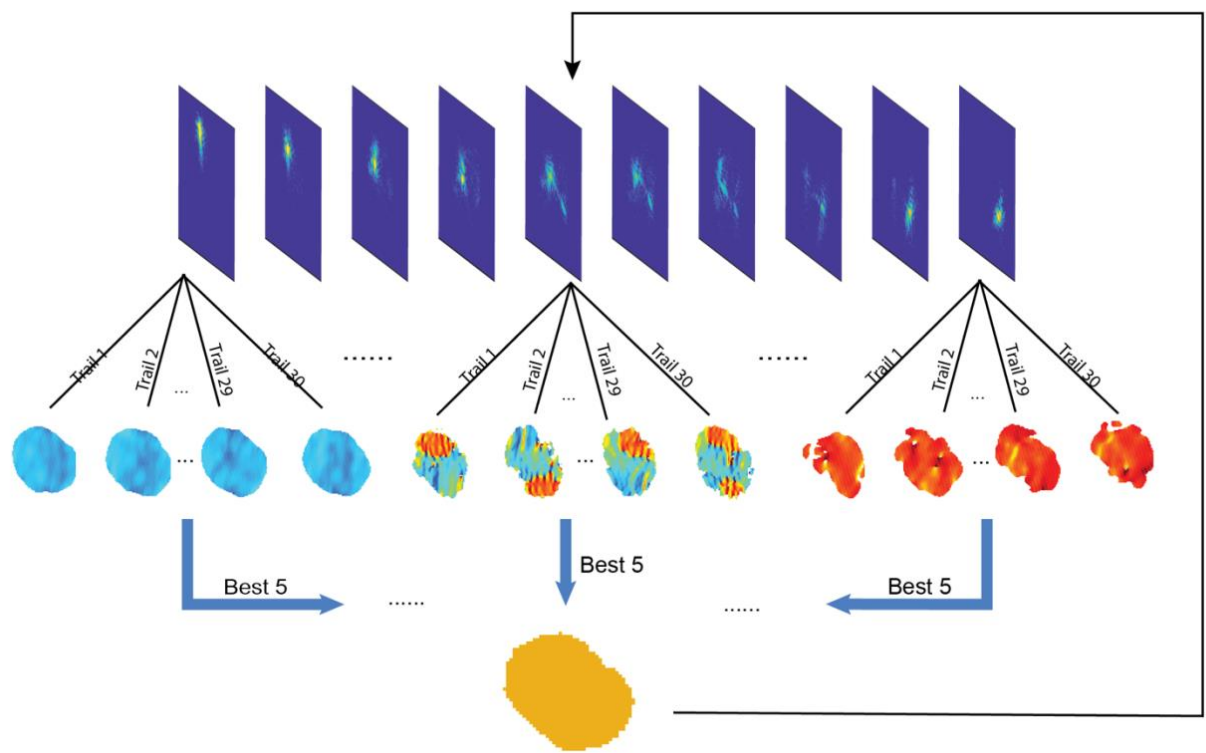

**Figure. S6. Illustration of the correlated phase retrieval algorithm.** The illustration shows one iteration of the algorithm. A set of scans (we used 10 scans) that include both single-peak and two-peak diffraction are reconstructed with alternating Error Reduction (ER)[2] and Relaxed Averaged Alternating Reflections (RAAR)[3] phase retrieval algorithms separately 30 times. Each individual reconstruction is called a trail. Out of the 30 trails for each scan, we select the best 5 trails. The support of the reconstruction for next iteration is then calculated by averaging the shape of the best 5 trails for all 10 scans. During the selection of the best 5 trails, the particle shape can appear inverted (two solutions (S(r) and S*(-r) are indistinguishable, where r is the coordinate and * denotes complex conjugate). To determine if the reconstruction is inverted, we use cross-correlate strain among different reconstructions.

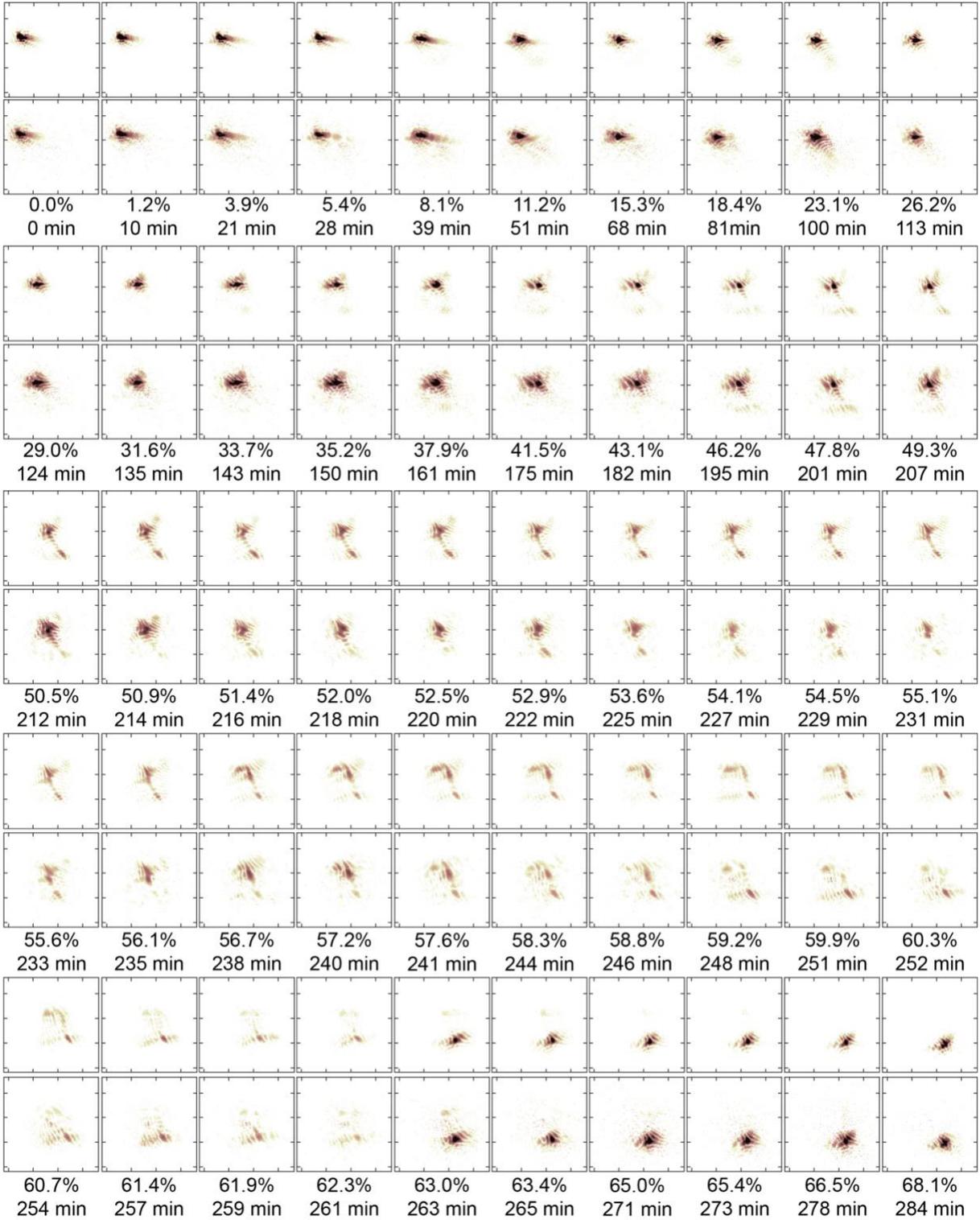

**Figure S7. A side-by-side comparison of the central theta slice from the measured diffraction (top) and the Fourier transform of reconstructions (bottom).** The close alignment between the reconstructed results and the diffraction data is indicative of successful phase retrieval. The false colors are identical to Fig. S4.

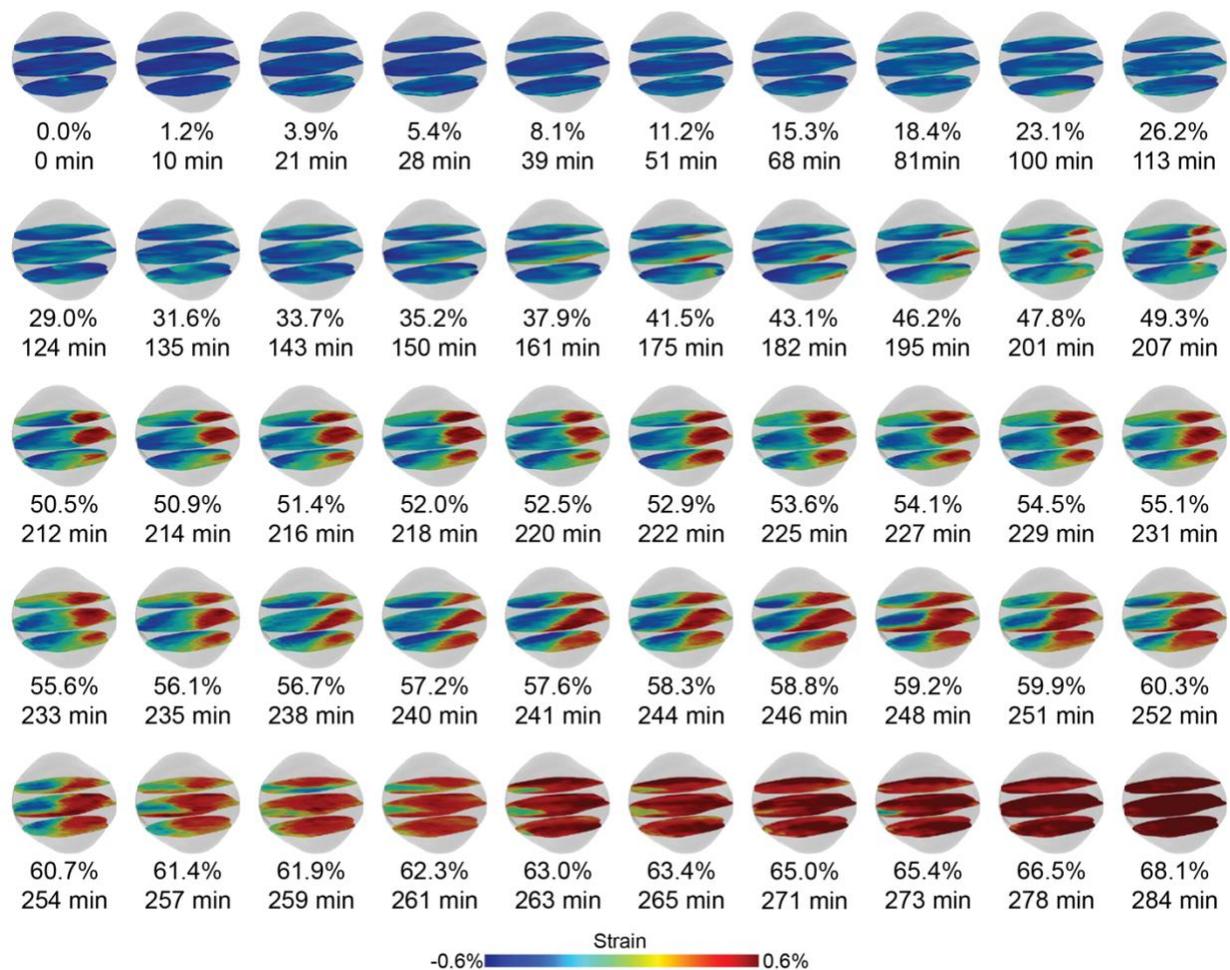

**Figure S8. Full set of the operando imaging of the structural phase transformation during discharge.** For each state of discharge, the chosen planes at the same position are imaged for the visualization of strain distribution inside the entire particle. The particle starts with a uniform negative strain (Li-poor phase) and ends the two-phase reaction with a uniform positive strain (Li-rich phase). The intermediate stages show that particle has both the red and blue phases, where the red phase grows at the expanse of blue phase, defining a nucleation and growth mechanism for the two-phase reaction. The particle is about 500 nm large.

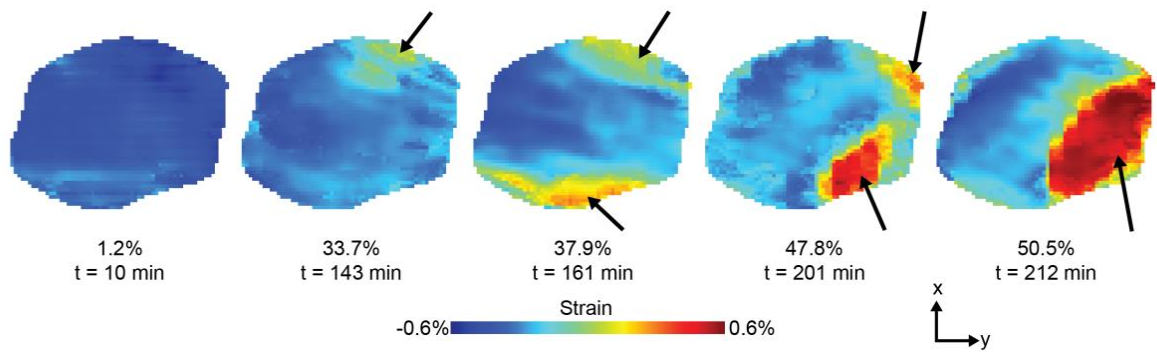

**Figure S9. 2D slice of the 3D strain distribution during the nucleation of the Li-rich phase.** The arrows indicate the nucleation sites, of the new Li-rich phase (red) inside the Li-poor phase (blue). When the Li-rich phase nucleates, it starts around the edge of the particle and can occur simultaneously at multiple locations. As the Li-rich phase continues to grow, the nucleated sites grow and, in the meantime, coalesce. Later, the nucleated sites merge for a total reduction of interface area.

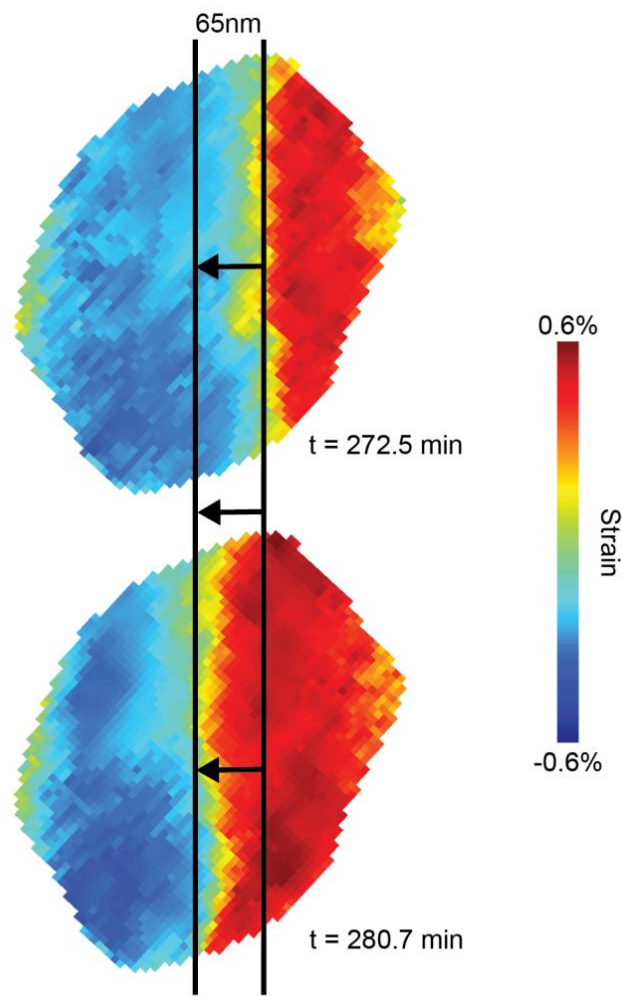

**Figure S10. Interface velocity measured during the operando imaging experiment.** Looking at specifically the region when a linear interface exists, we approximate the distance that the interface travels to be around 65 nm in 494 seconds. In three-dimensional diffusion in spinel $Li_xNi_{0.5}Mn_{1.5}O_4$, the movement of ions can be described with mean square displacement $<x^2> = 6Dt$, where $D$ is the diffusion coefficient and $t$ is the time. Using bulk diffusion coefficient of Li in $Li_xNi_{0.5}Mn_{1.5}O_4$, $2 \times 10^{-12}$ cm²/s [4], we calculate the average distance for the given time, $x = \sqrt{6Dt} = \sqrt{6 * 2 * 10^{-12} \text{ cm}^2/\text{s} * 494 \text{ s}} = 7.7 \times 10^{-5}$ cm $= 770$ nm. This is one order of magnitude larger than the 65 nm we observe. We conclude that at the discharge rate, the interface propagation is not limited by Li diffusion in the particle.

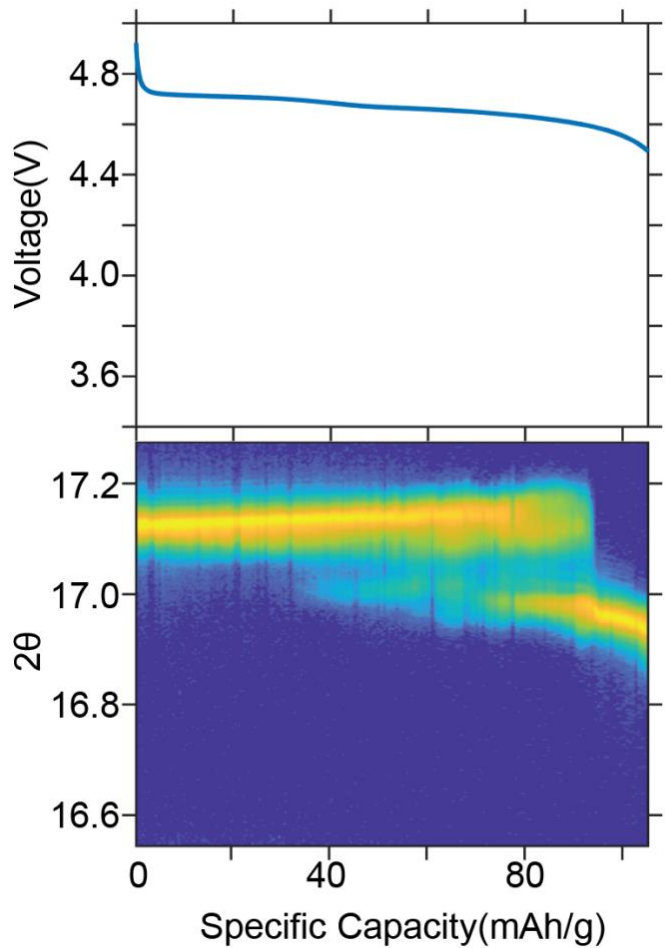

**Figure. S11. Additional electrochemical data of another cell (top) and diffraction data of another $Li_xNi_{0.5}Mn_{1.5}O_4$ particle that was discharged at C/10 (bottom).** Similar to the particle discussed above, this supplementary particle also displays an extended period of two-phase coexistence during the discharge, visible as the peak splitting in the diffraction. The two-phase region is again coincident with the voltage plateau in the electrochemical data.

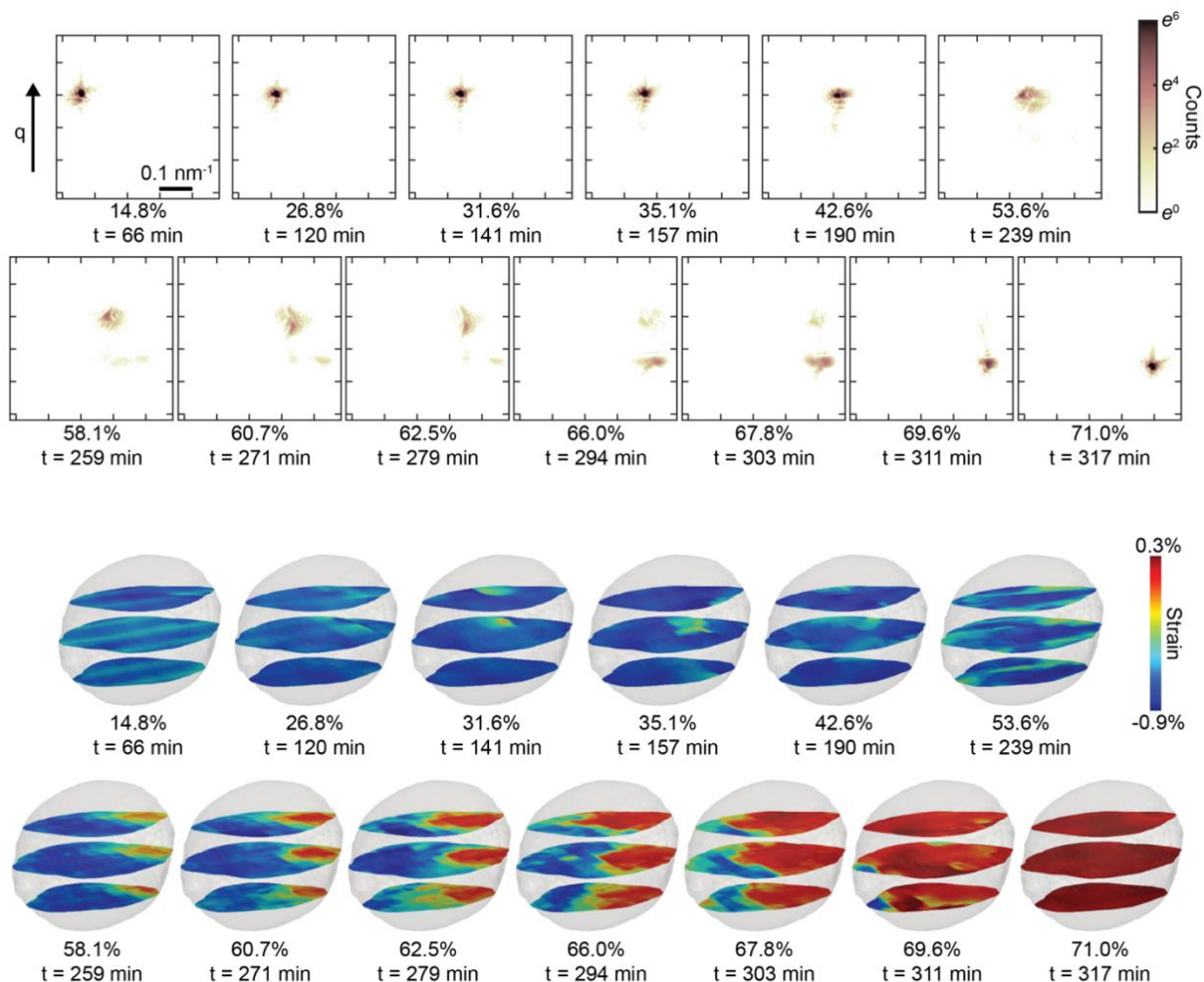

**Figure. S12. 2D diffraction slices and the corresponding inverted strain maps at three chosen planes for the supplementary particle during discharge at C/10.** Similar to the particle discussed above, this supplementary particle also shows a nucleation and growth regime where the nucleated Li-rich phase grows through interface propagation. The images are less smooth as they represent averages over less reconstruction runs.

**Microelasticity Modeling**

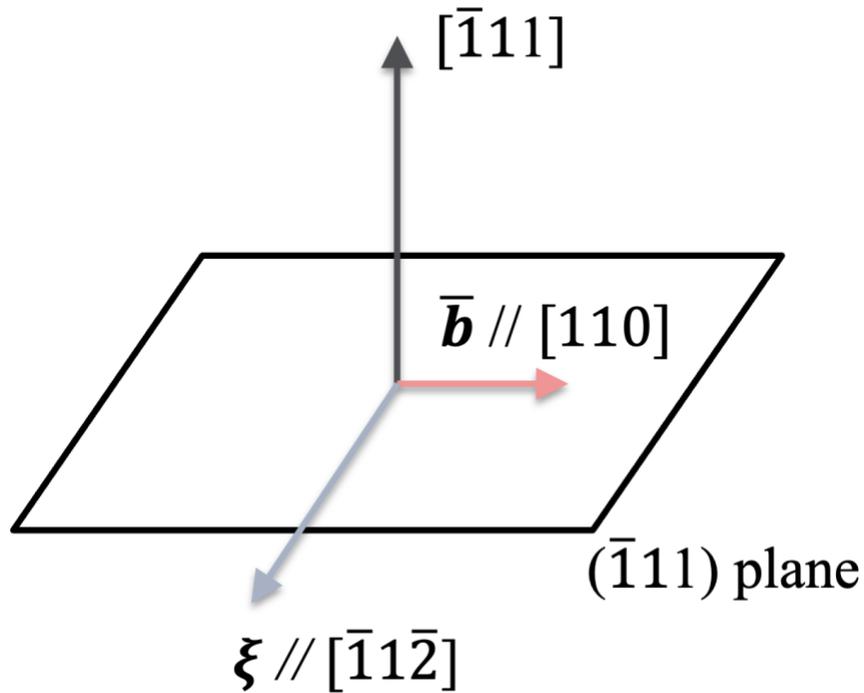

**Figure S13. Geometry of the dislocations used in microelasticity modelling.** During discharge, $Li_xNi_{0.5}Mn_{1.5}O_4$ transforms from a cubic-spinel to another cubic-spinel phase with a lattice mismatch of 1% (consistent with literature and estimated directly from the diffraction data through peak splitting). The coherency strain is the same along all three principal axes of the <100> family. We find the stress tensor by multiplying the strain tensor with the elastic stiffness tensor. The elastic stiffness tensor was adopted from a similar spinel material $LiTi_2O_4$ [5] and assumed equal for both phases. For the semi-coherent interface to reflect the coherency loss along [110], we rotate the strain matrix along the z direction by 45 degrees to align [110] on [100] and set it to be 0, then transformed it back to the previous coordinates.

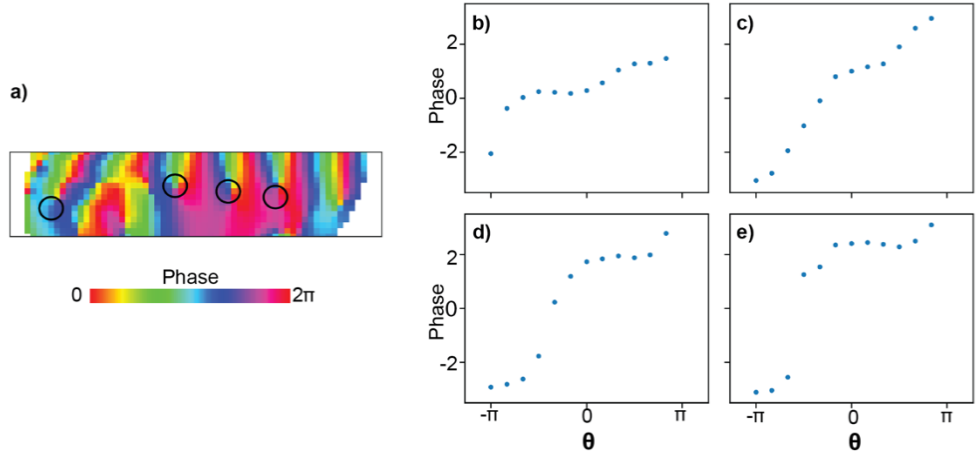

**Figure S14. Identifying singularities and the azimuthal scans around it indicating dislocations. a**) The enlarged phase map at 58.8% DoD same as Figure 4c. **b) - e)** The azimuthal scans around the singularities as circled in **a)**. All scans show an average phase jump around 5 radians.

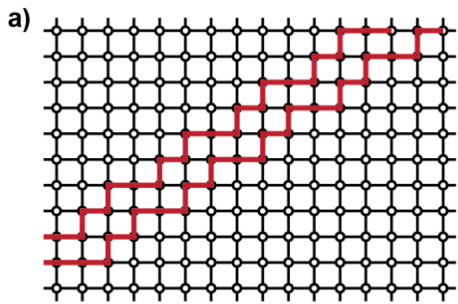

a)

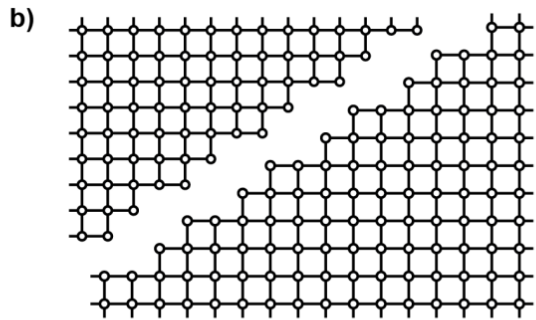

b)

**B**: lattice deformation

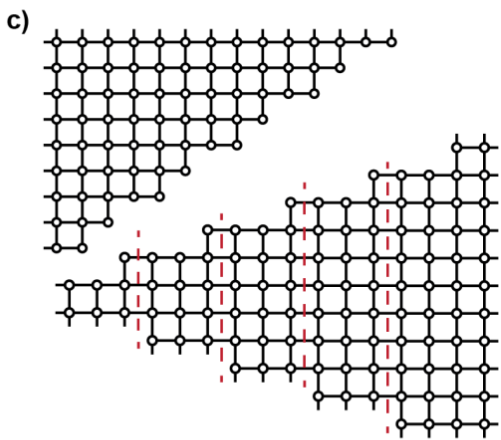

c)

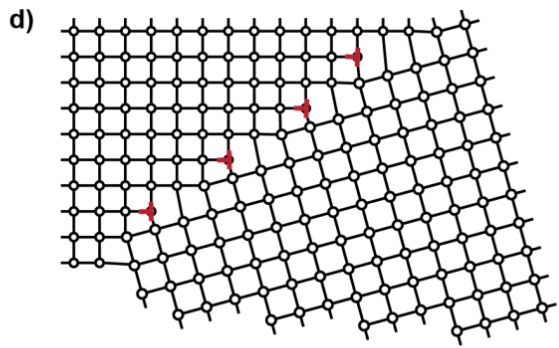

d)

**S**: lattice-invariant deformation

**R**: rotation

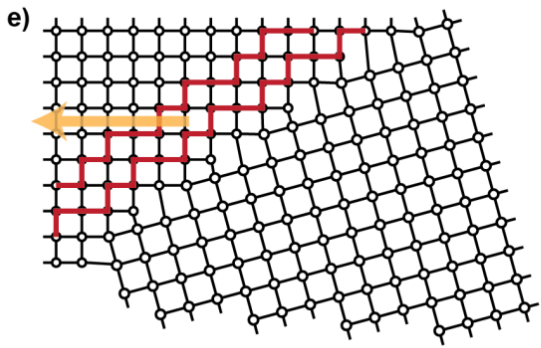

e)

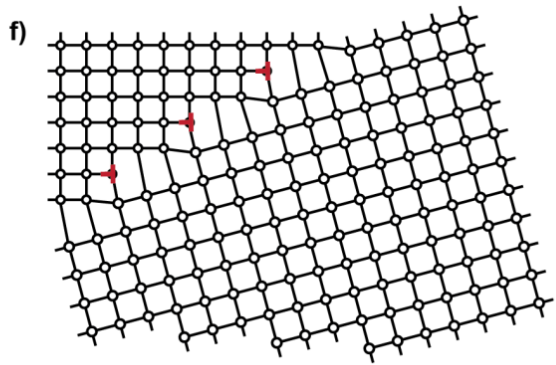

f)

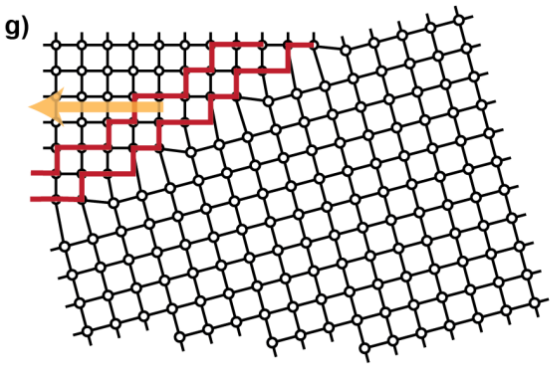

g)

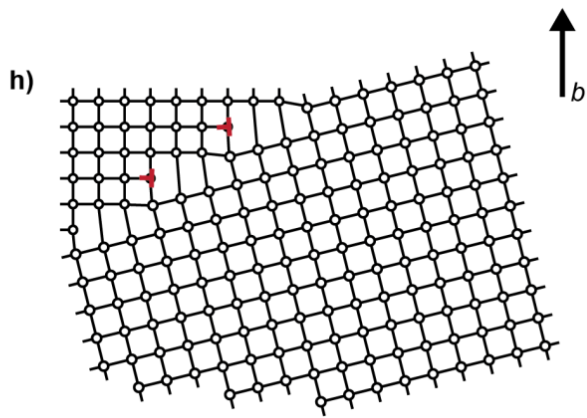

h)

**Figure S15. 2D illustration of the creation and propagation of a glissile interface.** The example shows a growth of a square structural phase with a larger lattice parameter, similar to growth of Li-rich $Li_xNi_{0.5}Mn_{1.5}O_4$ that has a cubic structure with a larger lattice parameter. The operation to minimize coherency strain at the interface includes lattice deformation ***B***, lattice-invariant shear deformation ***S***, and rotation ***R***.[6] The lattice-invariant deformation introduces slip at the interface and misfit dislocations at the interface, which move along slip planes during interface propagation. Consider snapshots during interface propagation, each snapshots shows an array of dislocations. During subsequent snapshots the dislocation array (each consisting of a different set of dislocations) moves perpendicular to the Burgers vector. This occurs via conservative motion where new dislocations enter the interface at the surface and exit the interface at the opposite surface.

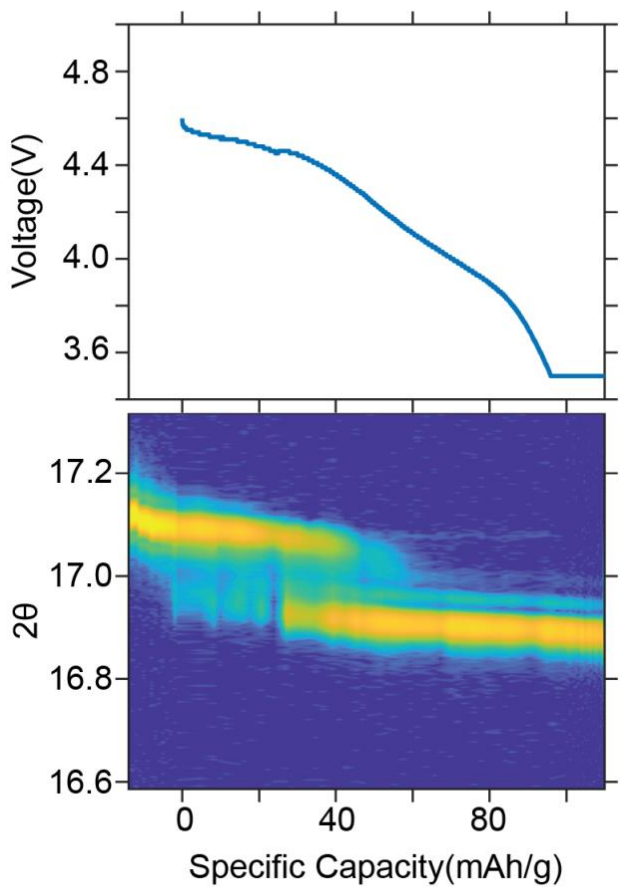

**Figure S16. Electrochemical data of the cell (top) and diffraction data of a single $Li_xNi_{0.5}Mn_{1.5}O_4$ particle within the cell (bottom) at C/2.** The two-phase coexistence is still present at a much higher discharge rate. This supports our hypothesis that the phase separation in $Li_xNi_{0.5}Mn_{1.5}O_4$ does not necessarily limit its kinetics as dislocations play an important role in reducing energy barriers for the reaction.

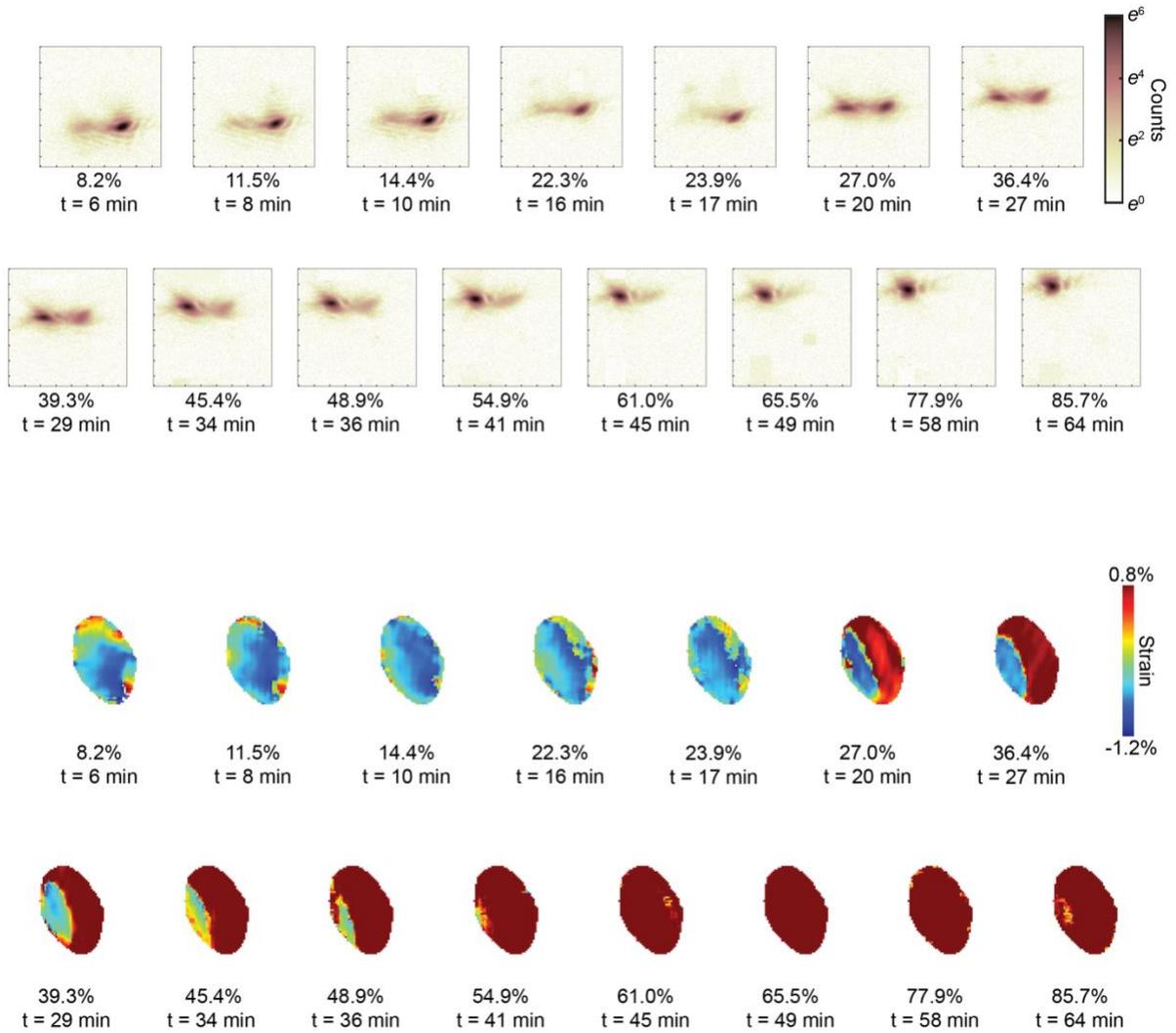

**Figure. S17. Operando imaging of the structural phase transformation during discharge at C/2.** Phase retrieval of the data shown in S16 and collected at a discharge rate of C/2. Akin to a slower discharge rate of C/10, nucleation and growth are visible.

**References**

1. Williams, G. J., Pfeifer, M. A., Vartanyants, I. A. & Robinson, I. K. Three-Dimensional Imaging of Microstructure in Au Nanocrystals. *Phys. Rev. Lett.* **90**, 4 (2003).

2. Fienup, J. R. Phase retrieval algorithms: a comparison. *Appl. Opt.* **21**, 2758 (1982).

3. Luke, D. R. Relaxed averaged alternating reflections for diffraction imaging. *Inverse Probl.* **21**, 37 (2004).

4. Kunduraci, M. & Amatucci, G. G. The effect of particle size and morphology on the rate capability of 4.7V LiMn1.5+δNi0.5−δO4 spinel lithium-ion battery cathodes. *Electrochimica Acta* **53**, 4193–4199 (2008).

5. Qi, Y., Hector, L. G., James, C. & Kim, K. J. Lithium Concentration Dependent Elastic Properties of Battery Electrode Materials from First Principles Calculations. *J. Electrochem. Soc.* **161**, F3010 (2014).

6. Balluffi, R. W., Allen, S. M. & Carter, W. C. *Kinetics of Materials*. (John Wiley & Sons, 2005).